\magnification1200

\vskip 2cm
\centerline {\bf Duality symmetries and phase space quantum theory}
\vskip 0.5cm

\centerline{Thomas  Curtright}
\vskip 0.5cm
\centerline{{\it Department of Physics, University of Miami,}}
\centerline{{\it Coral Gables, Florida 33146, USA}}
\centerline{}
\centerline{and}
\centerline{}
\centerline{Peter West}
\vskip 0.5cm
\centerline{{\it Mathematical Institute, University of Oxford,}}
\centerline{{\it Woodstock Road, Oxford, OX2 6GG, UK}}
\centerline{}
\centerline{{\it Department of Mathematics, King's College, London}}
\centerline{{\it The Strand, London WC2R 2LS, UK}}
\vskip 1cm
\centerline{curtright@miami.edu, peter.west540@gmail.com}
\vskip 2cm

\leftline{\sl Abstract}  
We argue that duality symmetries can be manifestly realised when theories with these symmetries are quantised using phase space quantum theory. In particular, using background fields and phase space quantum theory, we quantise the bosonic string and show that it has SO(26,26) symmetry, even when the string is not compactified on a torus.

\par 
\vskip2cm
\noindent

\vskip .5cm

\vfill
\eject
\medskip
{{\bf 1. Introduction}}
\medskip
The study of duality in fundamental physics began when Dirac [1] realised that if he introduced magnetic charges then 
the electric and magnetic fields of Maxwell could  be placed  on an equal  footing.   In this case Maxwell's theory possessed a  SO(2) duality symmetry. Similar duality symmetries have subsequently been discussed for many other theories.  The existence of such duality transformations can be traced back  to the construction of the irreducible representations of the Poincare group. Following the method of Wigner  [2] one finds a unique  irreducible representation for a particle of  a given spin and mass. However,    to construct a formulation that has manifest Poincare symmetry we must embed the irreducible  representation into a covariant representation of the Poincare group. That it is the same as the irreducible representation requires that  one employs gauge symmetry and enforces the field equation. 
\par
How to do the embedding is not unique as one can take different covariant representations.  This  leads to different formulations of the physical states. One can construct a theory that has two different embeddings, and so two  covariant representations,  but then one  should ensure that one still has  only the original physical states of the irreducible representation. To do this one must  impose a duality relation between the two different formulations. 
\par
To give a well known example, in eleven dimension supergravity  we can describe the non-gravitational bosonic states  by a three form  $A_{\mu_1\mu_2 \mu_3}$, or by a six form $A_{\mu_1\ldots \mu_6}$  and then we impose the duality relation  [3]
$$
F_{\mu_1\ldots \mu_4}={1\over 7!}\epsilon_{\mu_1\ldots \mu_4}{}^{\nu_1\ldots \nu_7} F_{\nu_1\ldots \nu_7}
\eqno(0.1)$$
where $F_{\mu_1\ldots \mu_4}=4 \partial_{[\mu_1} A_{\mu_2\ldots \mu_4 ]}$ and $F_{\mu_1\ldots \mu_7}=7 \partial_{[\mu_1} A_{\mu_2\ldots \mu_7 ]}$. 
Like in Maxwell's theory we can envisage  a symmetry  that transforms  the three form  gauge field into the six form gauge field, and visa-versa.  This is the case in E theory [4] and the transformations are part of the vast $E_{11}$ symmetry. In fact there are an infinite number of ways of embedding the irreducible representations of the Poincare group for any particle  into covariant representations and as a result on finds an infinite number of duality relations and so duality symmetries [4]. 
\par
Duality equations of the type of equation (0.1) involve an epsilon symbol and this relates the time derive acting on a gauge field to a spatial derivative acting on a gauge field. If we took a  Hamiltonian approach this equation would  relate, in field space,   a momentum to a coordinate, in other words, it relates two  conjugate variables. In  the quantum theory the momenta would obey a non-trivial commutator with the coordinates and taking a representation of this relation the momenta would be represented by derivatives with respect to the coordinates. As such the symmetry between the dual gauge fields that appears in the duality relation is no longer present. 
\par
The above discussion applies to the usual approach to quantum theory of Schr\"odinger, Heisenberg, Born and others. However, 
there is another approach developed by  Groenewold and Moyal in the 1940's [5,6] following earlier work of Weyl [7], Wigner [8] and  von Neumann [9]. For a review, and further references,  see  [10] and [11]. Although this approach to quantum theory  is equivalent to the familiar approach,  it is very different in that there are no operators. It involves  functionals of coordinates of momenta which  are multiplied according to a star product which is non-commutative. Using this approach to quantum theory one keeps the coordinates and the momenta on an equal footing  and so preserves the  way they appear  in the duality relations. As a result one finds a formulation that can preserve duality symmetry. 
\par
T duality is one  of the most remarkable symmetries of string theory [12]. This  SO(D,D) symmetry is present when a closed  string is compactified on a torus that has D dimensions. Crucial to this symmetry is the ability of the string to wrap around the circles of the tori. It is also present if there are certain isometries of the background spacetime. For a review,  and many references,  see  [12]. Partly inspired by T duality  a  formulation of the bosonic string world sheet action in the background of the massless field of the string that has an SO(D,D) symmetry has been constructed [13,14].  A Hamiltonian formulation  to this action, which we will extensively use,  was given in reference [15]. This was achieved by taking the string to have the  coordinates $x^\mu(\sigma) $ and  $y_\mu(\sigma) $ which belong to the vector representation of SO(D,D). This formulation did not require any compactification. 
\par
Taking a Hamiltonian approach to formulate the quantum theory one finds that    $x^\mu(\sigma) $ and  $y_\mu(\sigma) $ are conjugate variables and,  under the usual rules of quantum theory,  they obey a commutator whose representation takes one of the coordinates to be the derivative of the other. This choice breaks the SO(D,D) symmetry and proceeding in this way one finds the usual quantum string. 
\par
In section one we will give a brief review of phase space quantum theory that emphasises that it is a complete independent  approach to quantum theory  in  its own right. In section two we quantise the free spin zero particle in two dimensions in its dual formulation in order to illustrate how the phase space quantum theory works in a simple field theory example. In section three we will  quantise the above mentioned world sheet action of the bosonic string  in a background using phase space quantum theory. In this approach the state functional   depends on both $x^\mu(\sigma) $ and  $y_\mu(\sigma) $ and  the SO(D,D) symmetry is preserved.  Finally,  in section four we discuss the consequences of our results for future work. 
 

\medskip
{\bf 1. Brief review of quantum mechanics in phase-space}
\medskip
Since phase-space quantum theory is not well known we will give a very short review in this section. However, 
given the counter intuitive nature of this approach to quantum theory we will begin by recalling  some very well known features of the usual formulation of quantum theory  of Heiseinberg, Schr\" odinger, Dirac, Born  and others. In this way the reader can clearly see the differences compared to the phase-space formulation. 
\par
The formulation of the quantum theory often begins with the Hamiltonian formulation of the classical theory in which the momenta conjugate to the coordinates are identified and required to satisfy the Poisson bracket 
$$
\{ x^\mu , p_\nu \}= \delta ^\mu_\nu
\eqno(1.1)$$
In the usual formulation of quantum theory one takes all the measurable variables to be operators, which we  denoted with  a hat, and  one follows Dirac and imposes, for example, the relation 
$$
  [\hat x^\mu , \hat p_\nu ]=i\hbar \delta ^\mu_\nu
\eqno(1.2)$$
The next step is to take a representation of this relation which is most often taken to be 
$$
\hat x ^\mu= x^\mu , \ \hat p_\mu= -  i\hbar {\partial\over \partial x^\mu}
\eqno(1.3)$$
All the information about the system under study is encoded, in Dirac notation, by  $|\psi >$ which in this representation is given by $\psi (x^\mu )= <x |\psi >$  and so it depends on the coordinates of spacetime. One can take other representations of the commutator,  for example, in  the  momentum space one takes 
$\hat x ^\mu=  i\hbar {\partial\over \partial p_\mu} , \ \hat p_\mu = p_\mu$. However, whatever choice one makes one treats the coordinates and momenta in  different ways. 
\par
The measurement of a observable  $A$, with orthonormal eigenfunctions $  u_n (x)$,  is one of its eigenfunctions, denoted by  $\lambda_n$. As the eigenfunctions form a complete set we can express the state function in terms of the eigenfunctions as 
$$
 \psi (x )=\sum_n c_n  u_n (x)
\eqno(1.4)$$
The probability of a measurement of $A$ getting $u_n$ in the measurement is $c_n c_n^\star $. In Dirac notation we write 
$|\psi>= \sum_n c_n  |u_n >$ and so $c_n = <u_n |\psi>$ as $<u_n |u_m>=\delta _{nm}$. The average value of the measurement is given by $<A>= <\psi |\hat A |\psi>$. 
\par
An equivalent formulation of the usual formulation of quantum theory is given in terms of the density matrix $\hat \rho$ which is given for the above system by 
$$
\hat \rho = |\psi ><\psi |
\eqno(1.4)$$
In terms of the above eigenfunctions we can express $\hat \rho$ as 
$$
\hat \rho = \sum_{n, m} c_n c_m^\star |u_n> <u_m |\equiv  \sum_{n, m} \rho_{nm}  |u_n> <u_m |
\eqno(1.5)$$
where $  \rho_{nm}\equiv  c_nc_m^\star  = < u_n | \hat \rho | u_m >$. The probability of getting the result $\lambda_n$ in a measurement of $A$ is $\rho_{nn}$ while the average value of the measurement is given by 
$$
<A>= \sum_{n , m} <\psi | u_n ><u_n | A | u_m ><u_m| \psi >=\sum_{n , m}  \rho_{mn} <u_n | A | u_m > = Tr ( \rho A)
\eqno(1.6)$$
where we have used the fact that the eigenfunctions form a complete orthonormal set and so $ \sum_{n }| u_n ><u_n | =I$. 
Using the Schrodinger equation one readily finds the 
$$
i\hbar {\partial \hat \rho \over \partial t}= [\hat H , \hat \rho  ]
\eqno(1.7)$$
where $\hat H$ is the Hamiltonian operator. 
\par
In this paper we will only consider what is called pure states for which $\hat \rho$ can be written in the form of equation (1.4). We note that in this case $\hat \rho^2=\hat \rho$, $ \rho _{nn}\ge 0$  and  $\sum_n \rho_{nn}= 1$ as the probabilities must add to one. 
\par
The phase space formulation of quantum theory was first fully understood by Hilbrand Groenewold and Jos\'{e} Moyal in the 1940's
[5,6]. It takes a very different, and counter intuitive approach. As this  is a complete formulation of quantum theory  in its own right we will first explain it without reference to the usual formulation and then show its equivalence. 
\par
All the possible information about a system is encoded in a state function which which is a function of {\bf both} $x^\mu$ and $p_\mu$, that is,  $f(x^\mu , p_\mu )$! Given the Heisenberg uncertainty principle  it was not surprising that some of the pioneers of quantum theory, including Dirac, expressed their doubts as to whether quantum theory could be formulated in this way. In this formulation we do not have any operators and the measurable  observables  are also just ordinary functions of $x^\mu$ and $p_\mu$. What is different to the classical theory is the way one multiplies  quantities, it is not just straightforward usual multiplication,  but uses  what is called the star product. 
\par
For simplicity we will restrict our discussion to just one dimension. Given two state functions $f(x,p)$ and $g(x,p)$ the  star product is given by [5]
$$
f\star g \equiv f\left(  x,p\right)  \;\exp\left(  \overleftarrow{\partial_{x}
}\, {i\hbar\over 2}\,\overrightarrow{\partial_{p}}-\overleftarrow{\partial
_{p}}\,{i\hbar\over 2}\,\overrightarrow{\partial_{x}}\right)  \;g\left(
x,p\right) ,\ \ {\rm or} \ \ \star \equiv \exp\left(  \overleftarrow{\partial_{x}
}\, {i\hbar\over 2}\,\overrightarrow{\partial_{p}}-\overleftarrow{\partial
_{p}}\,{i\hbar\over 2}\,\overrightarrow{\partial_{x}}\right)  
\eqno(1.8)$$
The latter exponentiated derivatives are implemented as directed shift
operators  and so we may write the above as 
$$
f\star g   =f\left(  x+{1\over 2}i\hbar\,\overrightarrow{\partial_{p}
}\;,\;p-{1\over 2}i\hbar\,\overrightarrow{\partial_{x}}\right)  \;g\left(
x,p\right)  =f\left(  x,p\right)  \;g\left(  x-{1\over 2}\hbar
\overleftarrow{\partial_{p}}\;,\;p+{1\over 2}i\hbar\,\overleftarrow{\partial
_{x}}\right) 
$$
$$
  =f\left(  x+{1\over 2}\hbar\,\overrightarrow{\partial_{p}}\;,\;p\right)
\;g\left(  x-{1\over 2}\hbar\overleftarrow{\partial_{p}}\;,\;p\right)
=f\left(  x,p-{1\over 2}\hbar\,\overrightarrow{\partial_{x}}\right)
\;g\left(  x,p+{1\over 2}\hbar\,\overleftarrow{\partial_{x}}\right)
\eqno(1.9)$$
The star product can also be formulated in an integral form 
$$
f\star g=\int {dx_{1}dp_{1}\over 2\pi\left(  \hbar/2\right)  }\int
{dx_{2}dp_{2}\over 2\pi\left(  \hbar/2\right)  }\;f\left(  x+x_{1},p+p_{1}\right)
\;g\left(  x+x_{2},p+p_{2}\right)  \;\exp\left(  {i\over \hbar/2}\left(
x_{1}p_{2}-x_{2}p_{1}\right)  \right)
\eqno(1.10)$$
In the exponential we find the area of the parallelogram in phase space with vectors $(x_1,x_2)$ and $(p_1 ,p_2)$ divided by Planck's constant. We note that taking the complex conjugate of the star product exchanges the order, that is, $(f\star g)^\star = g\star f$.
\par
As we now show   the star product is non-commutativity.  These properties of the star product can be established by acting on arbitrary exponential functions linear in $x$ and $p$.  We find that 
$$
e^{ax+bp}\star e^{Ax+Bp}=e^{\left(  a+A\right)  x+\left(  b+B\right)
p}\;e^{\left(  aB-bA\right)  i\hbar/2}\ ,\
$$
$$
  e^{Ax+Bp}\star e^{ax+bp} =e^{\left(  a+A\right)  x+\left(  b+B\right)  p}\;e^{\left(  Ab-Ba\right)
i\hbar/2}
\eqno(1.11)$$
and therefore $e^{ax+bp}\star e^{Ax+Bp}\neq e^{Ax+Bp}\star e^{ax+bp}$. 
\par
However,  the star product  is associative 
$$
\left(  e^{ax+bp}\star e^{Ax+Bp}\right)  \star e^{\alpha x+\beta p}=e^{ax+bp}\star\left(  e^{Ax+Bp}\star e^{\alpha x+\beta p}\right)
$$
$$
=e^{\left(  a+A+\alpha\right)  x+\left(  b+B+\beta\right)  p}\;e^{\left(
aB-bA+a\beta {-}b\alpha+A\beta {-}B\alpha\right)  i\hbar/2}
\eqno(1.12)$$
These properties hold for all functions that admit Fourier,  or Laplace,  transform representations, and for all
 polynomials of $x$ and $p$. 
\par
The Moyal bracket is essentially just the
antisymmetric part of a star product [6], 
$$
\left[  f,g\right]  _{\star}\equiv f\star g-g\star f\ .
\eqno(1.13)$$
We find the consoling result   $\left[ x,p\right]  _{\star}=i\hbar$. Hence in the transition from the classical theory the Poisson bracket $ \{  f,g  \}$ is replaced by  $ {1\over i\hbar}\left[  f,g\right]  _{\star}$.
\par
The  star product,  as given above, obeys the Lone Star Lemma that applies to integrations of star  products over
all of phase-space.  Namely, one may use integrations by parts to supplant one, but generally only one,  star product in an integrand by an ordinary product as follows
$$
\int f\star g\ dxdp=\int f\,g\ dxdp=\int g\,f\ dxdp\,=\int g\star f\ dxdp
\eqno(1.14)$$
\par
The analogue of the Schr\"odinger equation concerns the time dependence of $f(x,y)$ which  is given by 
$$
i\hbar{\partial f\over \partial t} = [H , f  ]_\star
\eqno(1.15)$$
Given an observable $A$, which is a function of $x$ and $p$,  its  eigenfunctions $u_n (x, p) $ are given by 
$$
A\star u_n =\lambda_n u_n = u_n\star A 
\eqno(1.16)$$
\par
We will take the state functions $f(x,y)$ to be real. As we will discuss below this is the same as assuming,  in the usual formulation of quantum theory,  that we are considering only  pure states. As such we take  the eigenfunctions $u_n$  real and,  of course, the eigenvalues $\lambda_n$ are also  real. Taking the complex conjugate of the first of the above equations, and using the relation  $(A\star u_n )^\star= u_n \star A$,  we find the second equation. Taking the sum and difference of the two equations we find the real and imaginary parts of the equation. 
\par
The average value of a measurement of the observable $A(x,p)$ for a system with state function $f(x,p)$  is given by integrating over their product, namely 
$$
\int dx dp A(x,p)  f(x,p) 
\eqno(1.17)$$
\par
When measuring an observable $A$, the probability of finding the result $\lambda_n$ is 
$$
   {2\pi \hbar  } \int dxdp \ v_n(x,p)  f(x,p) 
\eqno(1.18)$$
As such the probability density  of finding the  particle at position $x$ is given by 
$$
    \int  dp f\left(  x,p\right)   
\eqno(1.19)$$
as the eigenfunction is $v(y,p) ={1\over 2\pi \hbar} \delta (y-x) $.  This is provided we normalise the state function by taking $\int dxdp f\left(  x,p\right)   =1$.
Similarly the  probability density  of finding the  particle with momentum  $p$ is given by 
$$
   \int  dx f\left(  x,p\right)  
\eqno(1.20)$$ 
\par
We will now explain how  phase space quantum theory is equivalent to the usual quantum theory. This can be seen by considering the Wigner function [9] which is given by 
$$
f\left(  x,p\right)  ={1\over \pi\hbar}\int dy\,\psi\left(  x+y\right)
\,\psi^{\ast}\left(  x-y\right)  \,e^{-2ipy/\hbar}\ ,
\eqno(1.21)$$
The inverse transformations being 
$$
 \psi\left(
x+y\right)  \,\psi^{\ast}\left(  x-y\right)  =\int dp\,f\left(  x,p\right)
\,e^{2ipy/\hbar} 
\eqno(1.22)$$
where as usual $\psi (x) = <x  | \psi >$ and  $\psi ^\star (y) = <\psi | y>$. What we have just called the Wigner function is actually the  state function $f(x,p)$ in phase space quantum theory. Usually discussions on phase space quantum theory do not use the term state function and just refer to the Wigner function. As we will comment below,  in order to recover the standard form of quantum theory from the above formulation of phase space quantum theory the above choice of Wigner function  requires  the definition of the star  product of equations (1.8) to (1.9). There are other possible choices. 
\par
These formulae can be written in terms of the density matrix of equation (1.4)  as 
$$
f\left(  x,p\right)  ={1\over \pi\hbar}\int    dy  \left\langle x+y\right\vert
\,{\hat \rho}{\,}\left\vert x-y\right\rangle \,e^{-2ipy/\hbar
} \ \ 
\eqno(1.23)$$
where $x$, $y$, and $p$ are {\bf not} operators. We note from equation (2.21) that  $f\left(  x,p\right)  $ is manifestly real. However, this equation  only follows from equation (1.23) if $\hat \rho$ is of the form of equation (1.4), that is, it is the density matrix corresponding to  a pure state. 
\par
It is instructive to compute a state function from a wavefunction in the usual quantum theory. The normalised eigenfunction with momentum $q$ in the usually quantum theory is ${1\over \sqrt {2\pi\hbar} }e^{{iqx\over \hbar}}$. Using equation (1.21) we find that the corresponding state function is given by $f(x,p)= {1\over 2\pi \hbar} \delta (q-p)$. Similarly starting from the eigenfunction of position  with eigenvalue $x_1$, that is $\delta (x-x_1)$, we find that the corresponding state function is that given below equation (1.19). 
\par
The Wigner function $f(x,p)$ and the density operator $\hat \rho $ contain the same information and can be expressed in terms of each other as follows 
$$
  \left\langle x+y\right\vert \hat \rho \left\vert
x-y\right\rangle =\int  dp f\left(  x,p\right)  \,e^{2ipy/\hbar} ,\ \  \hat \rho=2
\int dxdy dp\left\vert x+y\right\rangle f\left(  x,p\right)  \,e^{2ipy/\hbar
}   \,\,\left\langle x-y\right\vert \ 
\eqno(1.24)$$
While the first relation is obvious the second can be derived by substituting for $f (x,y) $ using  equation (1.23) and recognising  the relation $1=\int  dx \left\vert x\right\rangle \,\left\langle x\right\vert $, or more simply, substituting $\hat \rho$ in the first relation. 
\par
In  the usual formulation of quantum theory an observable is represented by an operator $\hat A$ which is itself  composed of the operators $\hat x$ and $\hat p$,   but in the phase space approach an observable  is represented by a real function of $x$ and $p$, that is, A(x,p). Weyl found a correspondence between these two quantities which is given by 
$$
\hat A (\hat x, \hat p)={1\over (2\pi)^2}\int d\tau d\sigma dx dp A(x,p) e^{i\tau (\hat p-p)+i\sigma (\hat x-x)}
\eqno(1.25)$$
with the inverse map 
$$
A(x,p) = \hbar \int dy <x+{\hbar \over 2}y |\hat A(\hat x,\hat p)|x- {\hbar \over 2}y> e^{-iyp}
\eqno(1.26)$$
This result can be proven by using the fact that $e^{i\tau \hat p}|x> = (x-\hbar \tau )|x>$ which follows if we use the relations  
$\hat xe^{i\tau \hat p}|x> = e^{i\tau \hat p}e^{-i\tau \hat p}\hat x e^{i\tau \hat p}|x>=   (x-\hbar \tau )|x>$ and $e^{i\tau (\hat p-p)+i\sigma (\hat x-x)}= e^{i\tau\sigma \hbar} e^{i\tau (\hat p-p)}e^{i\sigma (\hat x-x)}$.
\par
It is instructive to derive the average value of a measurement of an observable $A$ in the phase space quantum theory of equation (1.17) from the expression in the usual quantum theory, namely 
$$
<A>=<\psi |\hat A | \psi>={ 2}  \int dx dy <\psi | x+y><x+y | \hat A | x-y><x-y | \psi> 
$$
$$
= {2}  \int dx dy (\psi (x+y ) )^\star \psi(x-y) <x+y | \hat A | x-y> 
$$
$$
=  {2}  \int dx dy dp f(x,p) e^{-{2ipy\over \hbar}}<x+y | \hat A | x-y>=  \int dx dp\   f(x,p)  A(x,p)  
\eqno(1.27 )$$ 
where we have initially used equation (1.22) and then equation (1.26). 
One can also derive the probability of a given result of equation (1.20 ) starting from the result  in the usual formulation as follows 
$$
c_n^\star c_n = <v_n |\psi ><\psi |v_n>=    <v_n |\hat \rho |v_n>= 2\int dx dy dp (v_n(x+y))^\star v_n(x-y) e^{{2ipy\over \hbar}}
f(x,y)
$$
$$
=  2\int dx dy dp dq \ v_n(x,q) e^{{2i(p-q)y\over \hbar}}f(x,p)
= {2\pi \hbar   } \int dx dp \ v_n(x,p) f(x,p) 
\eqno(1.28)$$ 
In deriving this result we have  eliminated $\hat \rho$ using equation (1.24) and then used equation (1.22). 
\par
Using the correspondence of equations (1.25) and (1.25)  one can show that if the observables $A$ and $B$  in phase space quantum theory correspond to the operators $\hat A$  and $\hat B$ in usual quantum theory then $A\star B$ corresponds to 
$\hat A \hat B$. Of course this presumes a particular normal ordering. Indeed the star product we have given in equations (1.8-1.10) is the one required for the correspondence to the usual quantum theory found by taking the Wigner function to be that in equation (1.21). One can take alternatives to the  Wigner function in equation (1.21)  and then one has to take a different star product, see reference [11] for further discussions of this point. Since  the two formulations of quantum theory are equivalent  it follows that one can derive the Heisenberg uncertainty relation from the phase space approach, see reference [23] for a detailed discussion.  
\par
To illustrate phase space quantum theory,  and as they will be needed later  in this paper,  we will solve the harmonic oscillator. The energy eigenfunctions $v_n(x,p) $,  which have  energy eigenvalues $E_n$,  may be
obtained as real solutions of the eigenvalue equation (1.16) which in this case becomes
$$
H\star v=E\,\ v=v\star H
\eqno(1.29)$$
The Hamiltonian is given by the real function 
$$
H={1\over 2}\left(  p^{2}+x^{2}\right)= a^* a  = a^* \star a  + {\hbar\over 2}
\eqno(1.30)$$
where $a\equiv {1\over \sqrt{2}} (x+ip) $ and $a^*\equiv {1\over \sqrt{2}} (x-ip) $ 
For simplicity we take the mass and frequency of the harmonic oscillator to be  $m=1$ and $\omega=1$. 
\par
We define the vacuum by taking $a\star v_0=0 $ which using the star product of equation (1.9)  implies that 
$$
\{x+ {i\hbar \over 2}\partial_p) + i(p- {i\hbar \over 2}\partial_x)\}v_0=0
\eqno(1.31)$$
Taking real and imaginary parts we find that 
$$
(x+ {\hbar \over 2}\partial_x) v_0=0= (p+ {\hbar \over 2}\partial_p) v_0=0
\eqno(1.32)$$
and so $v_0= e^{-{1\over \hbar}(x^2+p^2)}$.  It is straightforward to verify that  $H\star v_0= {\hbar \over 2} v_0$ implying that the ground state energy $E_0={\hbar \over 2} $. 
\par
To find the other energy eigenstates we can proceed much like in usual quantum theory. One finds that $[a^* , a ]_\star = -\hbar $, $[H, a]_\star = -\hbar a$ and $[H, a^*]_\star = \hbar a^*$. Using these relations one finds that the energy eigenstates are given by 
$$
v_n= (a^*\star ) ^n \star v_0 \star  (a\star ) ^n
\eqno(1.33)$$ 
with energy $E_n= \hbar ({1 \over 2} +n)$. In this last equation $(a^*\star ) ^n \equiv a^*\star  (a^*\star ) ^{n-1}$  and similarly   $(\star a ) ^n \equiv \star a  (\star a ) ^{n-1}$. Finally we note for future use that the star product of equation (1.8) can be written in terms of $a^*, a$ variables as 
$$
\star   = e^{-{\hbar \over 2} ( \overleftarrow{\partial \over \partial {a^*}} \overrightarrow{\partial\over \partial {a}}
-  \overleftarrow{\partial\over \partial {a}} \overrightarrow{\partial\over \partial {a^*}} )} 
\eqno(1.34)$$ 


\medskip
{\bf 2 A spin zero in two dimensions} 
\medskip
As a warm up exercise for the bosonic string  we will first consider a free two dimensional scalar in its dual formulation. The equation of motion for a free spin zero particle $\varphi$, that   is,  $\partial_\mu \partial^\mu \varphi=0$ can be written in  terms of the  dual fields $\varphi$ and $\phi$  as 
$$
\partial_\mu \varphi= \epsilon_\mu{}^\nu \partial_\nu \phi \ \ {\rm or }\ \ \partial_\mu \phi= \epsilon_\mu{}^\nu \partial_\nu \varphi 
\eqno(2.1)$$
where $\epsilon ^{01}=1$ and so $\epsilon _{0}{}^{1}=-1$.
We adopt the Hamiltonian 
$$
H={1\over 2}  \int dx\{   \partial \varphi   \partial \varphi +    \partial \phi   \partial \phi \}
\eqno(2.2)$$
where $\partial = {\partial\over \partial x}$ and the Poisson bracket    
$$
\{\varphi (x) , \phi (y)\}= \epsilon (x-y)
\eqno(2.3)$$
where 
$$\epsilon (x) = \cases {{1\over 2}  \ \ \ , x >0, \cr  -{1\over 2} , x <0}
\eqno(2.4)$$
\par
The equation of motion (2.1) then follows by taking the Poisson bracket with the Hamiltonian, namely 
$$
 {\partial \varphi (x)\over \partial t}= \{ \varphi (x) , H \} \ \  {\rm or}\ \ {\partial \phi (x)\over \partial t}= \{ \phi (x) , H \} 
\eqno(2.5)$$
as equation (2.3) implies that $\{ \varphi (x) , \partial \phi (y)\}=- \delta (x-y)$. 
In this way of proceeding we do not, as is usually the case,  identify the coordinates and the conjugate momenta, instead we  simply adopt the Hamiltonian of equation (2.2)   the Poisson brackets and use these  to find the equations of motion. 
\par
We will now quantise this theory using phase space  quantum field theory. In this theory the  state functional $\Psi$ which is a functional, rather than an operator,  of the dual fields $\varphi$ and $\phi$, that is $\Psi (\varphi (x), \phi (x))$. The poisson brackets of equation 
(2.3) become replaced by the relation 
$$
[\varphi (x) , \phi (y)]_\star\equiv \varphi (x) \star  \phi (y)- \phi (y) \star \varphi (x)=i \hbar  \epsilon (x-y)
\eqno(2.6)$$
where the  star product is given by 
$$
\star= exp{ {i \hbar  \over 2}\int dx dy \{ { \overleftarrow \delta \over \delta \varphi (x)} \epsilon (x-y) {\overrightarrow  \delta \over \delta \phi (y)} + 
{ \overleftarrow \delta \over \delta \phi (x)} \epsilon (x-y) { \overrightarrow \delta\over \delta \varphi (y)}\} }
\eqno(2.7)$$
\par
In the usual formulation of quantum theory,   equation (2.6) would be replaced by a  commutator  and we would take   a representation in which one of the variables would essentially be the derivative of the other.  In this step preference  would necessarily be given  to either $\varphi$ or  $\phi$. However, in the phase space approach to quantum theory we  instead take the wave functional to depend on both $\varphi$ and $\phi$, that is $\Psi (\phi(\sigma), \varphi(\sigma))$,  so treating the two fields in the same way. 
\par
Instead of being a operator, as is the case in  the usual approach,  the Hamiltonian is just given by the same expression of equation (2.2) as in the classical theory. The equation of motion for the state functional follows from equation (1.15) and is given by 
$$
-i\hbar {\partial \Psi (\phi(\sigma), \varphi(\sigma))\over \partial t}=[\Psi (\phi(\sigma), \varphi(\sigma)), H ]_*
\eqno(2.8)$$
\par
We can also consider the case that the spin zero is self-dual by more or less taking $\phi=\varphi$ in the above. In particular we have the classical equation of motion $\partial_\mu \varphi= \epsilon_\mu{}^\nu \partial_\nu \varphi$, the Poisson bracket $ \{\varphi (x) , \varphi (y)\}= \epsilon (x-y)$ and the Hamiltonian $H={1\over 2}  \int dx   \partial \varphi   \partial \varphi $. The star product is then given by $\star= exp{ {i \hbar  \over 2}\int dx dy \{ { \overleftarrow \delta \over \delta \varphi (x)} \epsilon (x-y) {\overrightarrow  \delta \over \delta \varphi (y)}\}}$. 
\par
Under a SO(1,1) transformation the fields change  as $\delta \varphi =-\lambda \varphi$ and $\delta \phi =\lambda \phi$. While this preserves   the equations of motion and the Poisson bracket it does  not leave the Hamiltonian invariant. However, it would be a symmetry of the theory were we to introduce a background gravity field  $G$ into the Hamiltonian, namely take 
$$
H={1\over 2}  \int dx\{  G \partial \varphi   \partial \varphi +  G^{-1}  \partial \phi   \partial \phi \}
\eqno(2.9)$$
and tge variation  $\delta G= 2\lambda G$. This symmetry is inherited by the theory when quantised using phase space quantum theory.


\medskip
{\bf 3 The SO(D,D) Bosonic string theory}
\medskip
\medskip
{\bf 3.1 The classical theory}
\medskip
String theory sweeps out a world sheet, described by $x^\mu (\tau, \sigma)$, according to the Nambu action. It was observed that when the string moved in a spacetime which contained a torus of dimension $D$ then it has T duality symmetry, that is, an SO(D,D) symmetry [12]. This has been shown to be a symmetry of  string theory to all orders of perturbation theory. 
 Encouraged by the existence of this symmetry,   the dynamics for the first quantised  bosonic string was reformulated in terms of coordinates $x^\mu (\tau, \sigma)$ and $y_\mu (\tau, \sigma)$ which transform as a vector under  SO(D,D) and so can be written as $X^N=(x^\mu , y_\mu )$ [13,14]. 
 \par
 We will now review   this development in a formulation which used the Hamiltonian approach [15]. The  Hamiltonian is given by 
$$
H={1\over 2} \int d \sigma (\epsilon \partial_1 X^M G_{MN}\partial_1 X^N+
\delta  \partial_1 X^M \Omega_{MN}\partial_1 X^N )
\eqno(3.1.1)$$
where $\epsilon$ and $\delta  $ are new fields, $G_{MN}$ and
$\Omega^{MN}=\Omega^{NM}$ are given by
$$
G_{MN}=\left(\matrix {g_{\mu\nu}&0\cr 0& (g^{-1})^{\mu\nu}\cr}\right)
,\quad
\Omega^{MN}=\left(\matrix {0&\delta_{\mu}^{\nu}\cr \delta^{\mu}_{\nu}
&0\cr
}\right)
\eqno(3.1.2)$$
and $\Omega_{MN}=(\Omega^{-1})_{MN}$. The field  $g_{\mu\nu}$ is the
background metric and, for simplicity,  we  have set the background two form and dilaton to
zero. ÊThe tensor $\Omega^{MN}$ is an SO(D,D) invariant tensor and $G_{MN}$ is an element of SO(D,D), that is, 
$G\Omega G=\Omega^{-1}$. 
\par
The Poisson bracket is  given by 
$$
\{X^N (\sigma ), X^M (\sigma')\}=\epsilon Ê(\sigma-\sigma')\Omega^{MN}
\eqno(3.1.3)$$ 
\par
The momenta conjugate to $\epsilon$ and $\delta$ are absent and requiring that their time derivative vanish we find the constraints 
$$
C_1\equiv\partial_1 X^M G_{MN}\partial_1 X^N=0
\eqno(3.1.4)$$
and
$$
C_2\equiv \partial_1 X^M \Omega_{MN}\partial_1
X^N=0
\eqno(3.1.5)$$
Their  Poisson brackets realise  the Virasoro algebra. 
\par
The equations of motion of the string follow, in the usual way,   by taking Poisson brackets with the Hamiltonian
$$
{\partial X^N (\sigma ) \over \partial t}= \{ X^N  (\sigma ) , H\}= \delta \Omega ^{MN}G_{NP}\partial_1 X^P+
\epsilon \partial_1 X^M
\eqno(3.1.6)$$
This does not look like  the SO(D,D) invariant equation of motion of reference [13],  but as explained in reference [15],  it can be rewritten as 
$$
\Omega_{MN}\epsilon^{\alpha\beta}\partial_\beta X^N= \sqrt {-\gamma
}\gamma^{\alpha\beta}G_{MN} \partial_\beta X^N
\eqno(3.1.7)$$
provided we identify $\epsilon$ and $\delta$ with the two dimensional world sheet metric $\gamma^{\alpha \beta}$, namely 
$$
\delta =\tilde \gamma_{00},\quad \epsilon= -{\tilde \gamma^{01}\over
\tilde \gamma^{00}}
\eqno(3.1.8)$$
where  $\tilde \gamma ^{\alpha\beta}=\sqrt {-\gamma }\gamma^{\alpha\beta}$ and  $\gamma =\det \gamma_{\alpha\beta}$. 
\par
The constraints of equations (3.1.4) and (3.1.5) do not look to be invariant under world sheet Lorentz symmetry,  but using the equation of motion they can be written as 
$$
\partial_\alpha X^M \Omega_{MN}\partial _\beta X^N=0
\eqno(3.1.9)$$
and 
$$
\partial_\alpha X^M G_{MN}\partial _\beta X^N=0
\eqno(3.1.10)$$
In fact the latter constraint can be derived from the first constraint using the equation of motion. 
Hence despite appearances,  the Hamiltonian approach given at the beginning of this section results in a theory that is invariant under   world sheet Poincare transformations.  If were to introduce the background metric, two form and dilation the theory would be  SO(D,D) invariant. We note that not only the coordinates $X^M$ of the string,  but also the background fields  transform under this symmetry.


\medskip
{\bf 3.2 Quantisation of the string according to  Shr\"odinger and Heisenberg}
\medskip
To quantise the above dual  formulation of the bosonic string  the coordinates become operators which we denote by $\hat X^N =(\hat x^\mu (\sigma ), \hat y_\nu (\sigma))$ and  we adopt the  commutators
$$
[\hat X^M (\sigma ),\hat X^N (\sigma'] =i\hbar \Omega^{MN}\epsilon
(\sigma-\sigma')
\quad {\rm or \ equivalently } \quad [\hat x^\mu (\sigma ),\hat y_\nu
(\sigma' ) ]= i\hbar \delta^\mu_\nu \epsilon Ê(\sigma-\sigma')
\eqno(3.2.1)$$
We observe that  $\hat x^\mu (\sigma )$ and $\hat y_\nu (\sigma )$ are conjugate variables and we cannot measure both of them simultaneously. To implement these commutation relations we can adopt the Schr\"odinger representation 
$$
\hat x^\mu (\sigma)=x^\mu (\sigma)\quad {\rm and} \quad \hat y_\mu
(\sigma)= -i\int ^\sigma d\sigma' {\delta \over \delta x^\mu (\sigma')}
\eqno(3.2.2)$$
In this case the wavefunctional depends on $x^\mu (\sigma)$.  
One could  also  take  the equivalent momentum representation. 
$$
\hat x^\mu (\sigma)=i\int ^\sigma d\sigma' {\delta \over \delta y_\mu
(\sigma')} \quad {\rm and} \quad \hat y_\mu (\sigma)= Êy_\mu (\sigma)
\eqno(3.2.3)$$
and then the wavefunctional is a functional of $y_\mu (\sigma)$. Either way the quantisation procedure gives preference to one of the coordinates and in doing so breaks the SO(D,D) symmetry. 
\par
Taking, for example,  the choice of equation (3.2.2) we arrive at the standard picture of the second quantised bosonic
string, the   wavefunctional $\Psi$ depends on $x^\mu (\sigma)$ and the  constraints of equations
(3.1.4) and (3.1.5), or equivalently (3.1.9) and (3.1.10),  are imposed as  conditions on the wavefunctional. 
\par
Introducing the usual oscillators $\alpha_n^\mu$ and $\bar \alpha_n^\mu$ one  finds that the quantum string has as  massless fields  the graviton, a scalar and an antisymmetric tensor field, the  tachyon  and an infinite number of massive particles, for a review see chapter three of reference [16]. Since the wavefunctional only depends on $x^\nu (\sigma )$ there is only one zero mode  position and so only one zero mode  momentum, more precisely $\alpha_0^\mu =\bar \alpha_0^\mu$. Hence the SO(D,D) symmetry has disappeared. 


\medskip
{\bf 3.3 Quantisation using Groenewold and  Moyal quantum theory. }
\medskip
In this approach we no longer have any operators and the state functional as well as  the dynamical variables are just   functions of the position and momenta.  However they are multiplied together using a non-commutative star product rather than usual multiplication. The Poisson bracket of equation (3.1.3) becomes 
$$
[ X^N (\sigma_1 ), X^M (\sigma_2 ]_\star \equiv X^N (\sigma_1 )\star X^M (\sigma_2)- X^M (\sigma_2 )\star X^N (\sigma_1)
=i\hbar \epsilon Ê(\sigma_1-\sigma_2)\Omega^{MN}
\eqno(3.3.1)$$
where 
$$\star= exp{\{ {i \hbar \over 2}\int d\sigma_1 d\sigma_2 \{ {\overleftarrow \delta \over \delta X^P (\sigma_1)} \epsilon (\sigma_1-\sigma_2) {\overrightarrow\delta\over \delta X^Q(\sigma_2)} \Omega^{PQ}}\}
\eqno(3.3.2)$$
The reader can verify that this does indeed lead to equation (3.3.1). 
\par
In contrast to the usual approach,  the  states of the theory are described by  the state functional $\Psi$ which depends on $X^N (\sigma) $,  that is of $x^\mu (\sigma)$ {\bf and } $y_\mu (\sigma)$,  even though $x^\mu(\sigma)$ and $y_\mu(\sigma)$ are conjugate variables. 
The classical  constraints  $C_1$ and $C_2$  of equations (3.1.4) and (3.1.5)  remain functions of $X^N(\sigma)$  in the quantum theory but they act on the state functional using the star product. 
\par
However,  taking $C_1\star \Psi = 0=C_2\star \Psi$ is inconsistent for the same reason as in the usual approach. To resolve this we  introduce the  oscillators first  used in the early days of string theory. 
To this end we change variables to 
$$
Z^\mu\equiv (-x^\mu +y^\mu) ,\ \ \bar Z^\mu\equiv  (x^\mu +y^\mu) 
\eqno(3.3.3)$$
The relations of equation (3.3.1) become 
$$
[Z^\mu (\sigma_1 ) , Z^\nu (\sigma_2 ) ]_\star = -2i\hbar \eta^{\mu\nu}  \epsilon (\sigma_1-\sigma_2) ,\ \ 
[\bar Z^\mu (\sigma_1 ) , \bar Z^\nu (\sigma_2 ) ]_\star = 2i\hbar \eta^{\mu\nu} \epsilon (\sigma_1-\sigma_2) 
$$
$$
[Z^\mu (\sigma_1 ) , \bar Z^\nu (\sigma_2 ) ]_\star =0
\eqno(3.3.4)$$
\par
Using the equations of motion of equations (3.3.7) we find, in the absence of background fields,  that $\dot Z^\mu= Z^{\mu\prime}$ and 
 $\dot { \bar Z}^\mu=-\bar  Z^{\mu\prime}$ where $ \dot Z^\mu\equiv  \partial_0 Z^\mu$ and  $ Z^{\mu\prime}\equiv  \partial_1 Z^\mu$.
As a result we can write $Z^\mu (\sigma ) $ and $\bar Z^\mu (\sigma ) $ in the form 
$$
Z^\mu(\tau, \sigma) = -{1\over \sqrt \pi} \{ q^\mu +\alpha^\mu _0 (\tau+\sigma)+ \sum _{n\not= 0} {e^{-in(\tau+\sigma)} \over  -in}\alpha^\mu _n \} , 
$$
$$
\bar Z^\mu(\tau, \sigma) = -{1\over \sqrt \pi} \{ \bar q^\mu +\alpha^\mu _0 (\sigma-\tau)+ \sum _{n\not= 0} {e^{-in(\tau-\sigma)} \over  in}\bar \alpha^\mu _n \} 
\eqno(3.3.5)$$
\par
Differentiating equation (3.3.4) we find that 
$$
[Z^{\mu\prime} (\sigma_1 ) , Z^{\nu\prime} (\sigma_2 ) ]_\star = +2i\hbar \eta^{\mu\nu}  \delta^\prime (\sigma_1-\sigma_2) ,\ \ 
[\bar Z^{\mu\prime} (\sigma_1 ) , \bar Z^{\nu\prime} (\sigma_2 ) ]_\star =- 2i\hbar \eta^{\mu\nu} \delta^\prime (\sigma_1-\sigma_2) 
\eqno(3.3.6)$$
Starting from equation (3.3.5),  and taking suitable integrals over $\sigma$ of equation (3.3.4), we find that the oscillators obey the relations 
$$
[\alpha_n^\mu , \alpha ^\nu_m]_\star = n\hbar \delta_{m+n, 0} \eta^{\mu\nu} ,\ \ 
[\bar \alpha_n^\mu , \bar \alpha ^\nu_m]_\star = -n\hbar \delta_{m+n, 0} \eta^{\mu\nu} ,\ \ [\alpha_n^\mu , \bar \alpha ^\nu_m]_\star = 0
\eqno(3.3.7)$$
and  that 
$$
[q^\mu , \alpha_0^\nu ]_\star = i\hbar\eta^{mu\nu} ,\ \ [\bar q^\mu ,\bar  \alpha_0^\nu ]_\star = -i\hbar\eta^{mu\nu} ,
\eqno(3.3.8)$$
All other commutation relations between $q^\mu$, $\alpha_0^\nu$, $\bar q^\mu$ and $ \bar  \alpha_0^\nu$ vanishing. 
\par
The star product corresponding to the above commutator relations are given by 
$$
\star= exp\{ {\hbar} \{{i\over 2}\eta^{\mu\nu}({\overleftarrow \partial\over \partial q^\mu }{\overrightarrow\partial\over \partial \alpha_0^\nu}- {\overleftarrow \partial\over \partial \alpha_0^\mu}
{\overrightarrow \partial\over \partial q^\nu }) 
+ \sum_{p, p\not= 0}p\eta^{\mu\nu}   {\overleftarrow \partial\over \partial \alpha_p ^\mu } {\overrightarrow\partial\over \partial \alpha_{-p} ^\nu }\}
- (q^\mu \to \bar q^\mu , \alpha_p^\mu \to \bar \alpha_p^\mu )\} \}
\eqno(3.3.9)$$
\par
In the  phase space quantum theory approach the  wavefuctional $\Psi$ depends on  the oscillators $\alpha_n^\mu$ and $\bar \alpha_n^\mu$,  but also of the zero mode coordinates $q^\mu$ and $\bar q^\mu$,  or alternatively $x^\mu$ and $y^\mu$. 
\par
As usual we take the Fourier transform of the constraints and define 
$$
L_n = {1\over 4} \int _{-\pi}^{\pi} d\sigma (C_1-C_2)e^{in\sigma} =  {1\over 4} \int _{-\pi}^{\pi} d\sigma \partial_1 Z^\mu  \partial_1 Z^\nu \eta_{\mu\nu}e^{in\sigma}= {1\over 2} \sum_p \alpha_p^\mu\alpha_{n-p}^\nu\eta_{\mu\nu}
\eqno(3.3.10)$$
$$
\bar L_n = {1\over 4} \int _{-\pi}^{\pi} d\sigma (C_1+C_2)e^{-in\sigma} =  {1\over 4} \int _{-\pi}^{\pi} d\sigma \partial_1 \bar Z^\mu  \partial_1 \bar Z^\nu \eta_{\mu\nu}e^{-in\sigma}= {1\over 2} \sum_p \bar \alpha_p^\mu\bar \alpha_{n-p}^\nu \eta_{\mu\nu}
\eqno(3.3.11)$$
For simplicity we have set the background metric to be the Minkowski metric. 
\par
The physical states of the string are given by  imposing  the constraints on the state functional, namely
$$
L_n \star \Psi=0 = \bar L_n \star \Psi ,\ n\ge 0 ,\ \ (L_0-1) \star \Psi=0 =( \bar L_0-1) \star \Psi 
\eqno(3.3.12)$$ 
 \par
 We can write the most general state functional as 
 $$
 \Psi= \phi (x^\mu , y_\mu,\alpha^\mu_0, \bar \alpha^\mu_0) \Psi_0+ \ldots +h_{\mu\nu} (x^\mu , y_\mu ,\alpha^\mu_0, \bar \alpha^\mu_0)  \alpha ^\mu_{-1}\star  \bar \alpha ^\nu_{-1}\star\Psi_0\star  \alpha ^\mu_{-1}\star  \bar \alpha ^\nu_{-1}\star   +\ldots 
 \eqno(3.3.13)$$
 where $\alpha_n^\mu \star \Psi_0=0= \bar \alpha_n^\mu \star \Psi_0 , n\ge1$. We then apply the constraints of equations (3.3.12) and (3.3.13).  The state functional depends on both $q^\mu$ and $\bar q^\mu$, or equivalently $x^\mu $ and $y_\mu$. Using the star production of equation (3.3.9) we find in particular  that 
 $$
( L_0+L_0) \star \Psi=\{- {1\over 4}({\partial\over \partial q^\mu}\eta^{\mu\nu}  {\partial\over \partial q^\nu} +
  {\partial\over \partial \bar q^\mu}\eta^{\mu\nu}  {\partial\over \partial \bar q^\nu})+N+\bar N\}\Psi
  $$
  $$
  =\{- {1\over 8}({\partial\over \partial x^\mu}\eta^{\mu\nu}  {\partial\over \partial x^\nu} +
  {\partial\over \partial y_\mu}\eta_{\mu\nu}  {\partial\over \partial y_\nu})+N+\bar N\}\Psi=0
\eqno(3.3.14)$$
 and 
 $$
( L_0-L_0) \star \Psi=\{- {1\over 4}({\partial\over \partial q^\mu}\eta^{\mu\nu}  {\partial\over \partial q^\nu} -
  {\partial\over \partial \bar q^\mu}\eta^{\mu\nu}) +  {\partial\over \partial \bar q^\nu}+N-\bar N\}\Psi
  $$
  $$
 = \{- {1\over 4}{\partial\over \partial x^\mu}  {\partial\over \partial y_\nu} ++N-\bar N\}\Psi=0
 \eqno(3.3.14)$$
 where $N$ and $\bar N$ are the levels numbers of the states. 
 We note that these conditions apply to the perturbative string
\par
 What is different in the phase space approach to quantum theory is that the state functional depends as usual on $\alpha_n^\mu$, 
 $\bar \alpha_n^\mu$ and $x^\mu$ but also on $y_\mu$. These belong to  vector representations of SO(D,D). Including the background massless fields of the string we find that the quantum string has an SO(D.D) symmetry even though it is not compactified in any way.


\medskip
{\bf 4 Discussion } 
\medskip
Let us recall the past developments concerning  SO(D,D) string symmetry. The  motion of the bosonic string is traditionally described by a world sheet action involving  the string coordinates  $x^\mu(\tau, \sigma)$ and  the background graviton, the two form and a dilaton fields. The action possess world sheet and spacetime diffeomorphism symmetry as well as  the gauge symmetry of the two form. However, if the string moves through a spacetime part of which is a torus of dimension D then the string has a SO(D,D) symmetry [12]. Essentially to the appearance of this symmetry is the wrapping of the string around the circles of the torus and this is not contained in the traditional account of string theory just mentioned.
\par
One can account of for this symmetry by constructing a quantum theory in which one splits  the string coordinates $x^\mu(\tau, \sigma)$ into left and right moving parts, $x_L^\mu(\tau, \sigma)$
and $x_R^\mu(\tau, \sigma)$  and then takes these to be completely independent of each other. A given torus compactification has  fixed background  graviton and  two form  fields and then  we only have an SO(D,D,Z) symmetry as only in this case do the particles contained in the string have the same masses. We can think of  $x_L^\mu(\tau, \sigma)$
and $x_R^\mu(\tau, \sigma)$  as belonging  to the vector representation of SO(D,D) and the background fields as 
 belonging  to the coset $SO(D,D) \over SO(D)\times SO(D)$.   The different tori  can be parameterised by $SO(D,D) \over O(D)\times O(D)$. We refer the reader to reference [12] for an extensive list of references. 
\par
Alternatively one can begin with the formulation of the bosonic string which has coordinates $x^\mu(\tau, \sigma)$ and $y_\mu(\tau, \sigma)$,  which belong to the vector representation of  SO(D,D), in the same background  fields [13,14]. Formulated in this way the action has an SO(D,D) symmetry even though there is no compactification. This construction was motivated by trying to give a world sheet account of the O(D,D,Z) symmetry found on compactification on a torus [14]. Another motivation was to recover from the string perspective the symmetries found in supergravity theories [13]. 
\par
However, the variables   $x^\mu(\tau, \sigma)$ and $y_\mu(\tau, \sigma)$ are conjugate variables and if we apply the usual rules of quantum theory to their non-trivial commutator one of the variables will be represented by a derivative of the other and the wavefunctional just depends on the other variable. In this way the SO(D,D) symmetry is no longer present.
\par
 In this paper we quantised the dual formulation of the bosonic string involving the coordinates $x^\mu(\tau, \sigma)$ and $y_\mu(\tau, \sigma)$ using phase space quantum theory. In this approach the state functional depends on both  $x^\mu(\tau, \sigma)$ and $y_\mu(\tau, \sigma)$ which belong to the vector representation of SO(D,D). We found that the quantum bosonic string in the background possesses a SO(D,D) symmetry even though it was not compactified. The zero mode coordinates of  
 $x^\mu(\tau, \sigma)$ and $y_\mu(\tau, \sigma)$ are not related to each other and as such the state functional depends on a spacetime that has $2D$ coordinates. Thus,  in effect,  the dimension of the spacetime has  doubled. 
 \par
 One could construct the interacting theory using the original vertex operator approach to string theory but instead using phase space quantum theory. In other words one would  introduce vertex operators constructed from $Z^\mu (\tau, \sigma )$ and   $\bar Z^\mu (\tau, \sigma )$. The tachyon vertex operators would be given by 
 $$
 V_T(z) \equiv e_*^{ik_\mu Z^\mu (z)} , \ \ {\rm and }\ \  \bar V_T(z)\equiv e_*^{i\bar k_\mu \bar Z^\mu (z)} 
 \eqno(4.1)$$
 where $z=e^{i\sigma}$ and $e_*^{f}\equiv 1+ f+{1\over 2} f\star f +\ldots $. While the  graviton vertex might be  given by 
 $$
 V^{\mu\nu}\equiv Z^\mu (z) \star \bar  Z^\nu (z) \star e_*^{ik_\rho Z^\rho (z)} e_*^{i\bar k_\rho \bar Z^\rho (z)} 
 \eqno(4.2)$$
 The particle scattering could be computed using the star product in correlations whose  the generic form is given by 
 $$
 \Psi_0\star V(z_1)\star \dots \star V(z_n)\star \dots  \star\Psi_0
  \eqno(4.3)$$
  \par
  The resulting interacting string theory would have a SO(D,D) symmetry and in particular the low energy effective action involving the graviton, two form and dilaton would also have this symmetry. 
  \par
  This should be essentially  consistent with Siegel theory [17,18,19] which contains  the metric, two form and dilaton which depend on coordinates that belong to the vector representation of O(D,D). While this theory is SO(D,D) invariant it also has a local gauge symmetry generalising diffeomorphism and gauge transformations. This theory emerges from E theory if one takes level zero in the decomposition corresponding to the  IIA theory [20]. The extension of Siegel theory to include the massless fields in the Ramond-Ramond sector of the superstring was found in [21], it is just the extension to include level one. More recently it has been shown that the non-linear realisation of the semi-direct product of the very extended algebra $D_{24}^{+++}$ with its vector representation contains the low effective action of the  bosonic strings and branes [24]. 
  \par
  The additional coordinates $y_\mu$ can be traced in E theory to the presence of the M2 brane  which leads to a coordinate $x_{\mu\nu}$  in eleven dimensions. The  coordinates  in E theory, including those of the usual spacetime,  can be thought of as arising from  the  moduli of solitonic object, such as the M2 brane. Indeed they are required to describe the dynamics of the solitons [22] and the very existence of  solitonic objects requires these  coordinates. Thus one can think of the enlargement of spacetime to include  of the additional coordinates  $y_\mu$ as just an example of the need to include additional coordinates that are required to account for solitons [22]. 
 \par
 We can think  of the calculations in this paper as an example of a more general phenomenon involving  duality symmetries between form fields which necessarily relate coordinates to momenta. Such a symmetry is not preserved by the usual approach to quantum theory as it treats these two quantities differently. However they are treated the same way if we use phase space quantum theory and so the duality symmetries are preserved. We hope to discuss other theories in the future and in particular the generalisation to the M2 brane.

\medskip
{\bf Acknowledgement}
\medskip
This work was initiated and began during the BASIC 25 workshop and we wish to thank our co-organiser Eduardo Guendelman. Peter West  also thanks the SFTC for support from Consolidated grant Pathways between Fundamental Physics and Phenomenology, 
ST/T000759/1.
\medskip
{\bf References}
\medskip
\item{[1]} P. Dirac, {\it Quantised singularities in the electromagnetic field} , Proc Roy. Soc. {\bf A33} 60. 
\item{[2]} E. Wigner, {\it On unitary representations of the inhomogeneous Lorentz group}. Annals of Mathematics. {\bf 40} 1 (1939) 149. 
\item{[3]} H. Nishino, {\it Alternative formulation for duality symmetric eleven-dimensional supergravity coupled to super M five-brane }, Mod.Phys.Lett.A 14 (1999) 977,  hep-th/9802009.
\item{[4]} P. West, {\sl $E_{11}$ and M Theory}, Class. Quant. Grav. {\bf 18 } (2001) 4443, {\tt hep-th/0104081};  {\sl $E_{11}$, SL(32) and Central Charges}, Phys. Lett. {\bf B 575} (2003) 333-342, {\tt hep-th/0307098}; for a review see P. West,  {\it  A brief review of E theory}, Proceedings of Abdus Salam's 90th  Birthday meeting, 25-28 January 2016, NTU, Singapore, Editors L. Brink, M. Duff and K. Phua, World Scientific Publishing and IJMPA, {\bf Vol 31}, No 26 (2016) 1630043,  arXiv:1609.06863.  
\item{[5]} H. Groenewold, {\it On the Principles of Elementary Quantum Mechanics},  Physica 12 (1946) 405-460.
\item{[6]}  J E Moyal, {\it Quantum Mechanics as a Statistical Theory},  Proc Cambridge Phil Soc 45 (1949) 99-124.
M Bartlett and J Moyal, {\it The Exact Transition Probabilities of Quantum-Mechanical Oscillators Calculated by the
Phase-Space Method},  Proc Camb Phil Soc 45 (1949) 545.
\item{[7]}H Weyl, {\it Quantum mechanics and group theory},  Z Phys 46 (1927) 1; The theory of groups and quantum mechanics, Dover, 1931. 
\item{[8]} J. von Neumann, {\it The uniqueness of the Schr\"odinger equation}, Math. Ann. {\bf 104} (1931) 570. 
\item{[9]} E Wigner, {\it Quantum Corrections for Thermodynamic Equilibrium}  Phys Rev 40 (1932) 749.
\item{[10]} D Fairlie, {\it The Formulation of Quantum Mechanics in Terms of Phase Space Functions}, 
Proc Camb Phil Soc 60 (1964) 581.
\item{[11]}C Zachos, D Fairlie, and T Curtright, {\it Quantum Mechanics in Phase Space}, World Scientific (2005).
\item{[12]}  T. Buscher, {\it A Symmetry of the String Background Field Equations},  Phys.Lett.B 194 (1987) 59-62; 
{\it Path Integral Derivation of Quantum Duality in Nonlinear Sigma Models},    Phys.Lett.B 201 (1988) 466-472; 
  for a review and further references see A. Giveon, M. Porrati and E. Rabinovici, {\it Target space duality in string theory}, Phys. Rept. 244, 77 (1994) [arXiv:hep-th/9401139].
\item{[13]} M. Duff, {\it Duality Rotations In String Theory}, Nucl.\ Phys.\ ÊB {\bf 335} (1990) 610; M. Duff and J. Lu,
{\sl Duality rotations in membrane theory}, ÊNucl. Phys. {\bf B347} (1990) 394.
\item{[14]} A. Tseytlin, {\it  Duality Symmetric Formulation of String World Sheet Dynamics},  Phys.Lett. {\bf B242} (1990) 163;
{\it Duality symmetric closed string theory and interacting chiral scalars}, Nucl. Phys. B350 ( 1991) 395
\item{[15]} P. West, {\it Generalised space-time and duality}, Phys.Lett. {\bf B693} (2010) 373, arXiv:1006.0893 
\item{[16]} P. West, {\it  Introduction to Strings and Branes}, (2012), Cambridge University Press. 
\item{[17]} Warren Siegel. {\it Superspace duality in low-energy superstrings}, Phys. Rev. D, 48 (1993) 2826.
\item{[18]} Warren Siegel. {\it Two vierbein formalism for string inspired axionic gravity}, Phys.Rev. D, 47 (1993) 5453, 
\item{[19]}  C. Hull and B. Zwiebach, {\it Double Field Theory},  JHEP {\bf 0909} (2009) 099, hep-th/0904.4664; 
O. Hohm, C. Hull and B. Zwiebach, {\it Background independent action for double field theory},  JHEP 07 (2010) 016, hep-th/1003.5027.
\item{[20]}  P. West, {\it E11, generalised space-time and IIA string theory},  Phys.Lett.B696 (2011) 403-409,
arXiv:1009.2624. 
\item{[21]}  A. Rocen and P. West, {\it  E11, generalised space-time and IIA string theory; the R-R sector}, in
Strings, Gauge fields and the Geometry behind:The Legacy of Maximilian Kreuzer edited by
Anton Rebhan, Ludmil Katzarkov, Johanna Knapp, Radoslav Rashkov, Emanuel Scheid,
World Scientific, 2013, arXiv:1012.2744.
\item{[22]} P. West,  {\it Spacetime and large local transformations},   Int.J.Mod.Phys.A 38 (2023) 08, 2350045, arXiv:2302.02199.
\item{[23]} T. Curtright and C. Zachos, {\it Negative Probability and Uncertainty Relations}, Mod. Phys. Lett. A16 (2001) 1381, https://arxiv.org/abs/hep-th/0105226. 
\item{[24]}  K.  Glennon and P. West, {\it K27 as a symmetry of closed bosonic strings and branes },  Int.J.Mod.Phys.A 40 (2025) 02, 2450155, arXiv:2409.08649  

\end